# Coherent energy exchange between vector soliton components in fiber lasers

**H. Zhang, D. Y. Tang\*, and L. M. Zhao**

School of Electrical and Electronic Engineering, Nanyang Technological University,

Singapore 639798

**N. Xiang**

Center for Optoelectronics, Department of Electrical and Computer Engineering,

National University of Singapore, Singapore 117576

\*: Corresponding author, edytang@ntu.edu.sg

We report on the experimental evidence of four wave mixing (FWM) between the two polarization components of a vector soliton formed in a passively mode-locked fiber laser. Extra spectral sidebands with out-of-phase intensity variation between the polarization resolved soliton spectra was firstly observed, which was identified to be caused by the energy exchange between the two soliton polarization components. Other features of the FWM spectral sidebands and the soliton internal FWM were also experimentally investigated and numerically confirmed.

Passive mode-locking of erbium-doped fiber lasers with a semiconductor saturable absorber mirror (SESAM) has been extensively investigated [1, 2]. In contrast to the nonlinear polarization rotation (NPR) mode-locking, mode-locking incorporating a SESAM does not require any polarization element inside the laser cavity, thereby under suitable condition of the cavity birefringence, vector solitons could be formed in the lasers [3]. Recently, it was reported that even the polarization-locked vector soliton (PLVS) could be formed in the mode-locked fiber lasers [4, 5]. Formation of a PLVS requires not only that the group velocities of the two orthogonal polarization components of a vector soliton are locked but also that their phase velocities are also locked. It is well known that through self-phase modulation (SPM) and cross-phase modulation (XPM), nonlinear interaction between the two polarization-modes of a fiber could result in group velocity locked vector solitons [6]. Although it was also pointed out that the four-wave-mixing (also called coherent energy exchange) coupling between the polarization components of a vector soliton could have contributed to the formation of the phase locked vector solitons [4, 5], so far no experimental evidence of soliton internal FWM has been presented. In this Letter we report on the experimental observation of FWM between the two orthogonal polarization components of a vector soliton formed in a fiber laser passively mode locked with a SESAM. Energy exchange between the two orthogonal polarization components of vector solitons was observed at specific frequencies on the soliton spectrum. However, our experimental results showed that the existence of FWM didn't guarantee formation of PLVS.

The fiber laser is illustrated in Fig.1. It has a ring cavity consisting of a piece of 4.6 m Erbium-doped fiber (EDF) with group velocity dispersion parameter 10 ps/km/nm and a total length of 5.4 m standard single mode fiber (SMF) with group velocity dispersion parameter 18 ps/km/nm. The cavity has a length of $4.6_{EDF}+5.4_{SMF}=10m$. Note that within one cavity round-trip the signal propagates twice in the SMF between the circulator and the SESAM. A circulator is used to force the unidirectional operation of the ring and simultaneously to incorporate the SESAM in the cavity. An intra cavity polarization controller is used to change the cavity's linear birefringence. The laser is pumped by a high power Fiber Raman Laser source (BWC-FL-1480-1) of wavelength 1480 nm. A 10% fiber coupler is used to output the signals. The laser operation is monitored with an optical spectrum analyzer (Ando AQ-6315B), a 26.5 GHz RF spectrum analyzer (Agilent E4407BESA-E SERIES) and a 350 MHz oscilloscope (Agilent 54641A) together with a 5 GHz photodetector. A commercial autocorrelator (Femtochrome FR-103MN) is used to measure the pulse width of the soliton pulses. The SESAM used is made based on a GaInNAs quantum well. The SESAM has a saturable absorption modulation depth of 5%, a saturation fluence of 90μJ/cm$^2$ and 10 ps relaxation time. The central absorption wavelength of the SESAM is at 1550nm.

Experimentally, it was noticed that after mode-locking multiple mode locked pulses were always initially formed in the cavity. Depending on the net cavity birefringence, they were either the group-velocity locked vector soliton, characterized by the rotation of soliton polarization state along the cavity, or the phase-locked vector soliton, characterized by the fixed polarization at the laser output. With multiple vector solitons in cavity, as a result of mutual soliton interaction complicated relative soliton movement or vector soliton bunches with random fixed soliton separations were observed. To exclude complications caused by soliton interactions, we have always reduced the number of soliton in cavity through carefully decreasing pump power so that only one or a few widely separated solitons exist in cavity.

Fig. 2 shows typical measured optical spectra of the PLVSs of the laser. The soliton feature of the mode-locked pulses is confirmed by the existence of soliton sidebands. As firstly reported by Cundiff et al [7], apart from the existence of the conventional Kelly soliton sidebands, on the vector soliton spectrum there exist also extra sets of spectral sidebands. Experimentally it was noticed that different from the Kelly sidebands whose positions are almost independent of the laser operation conditions, such as the pump strength and polarization controller orientation change (linear cavity birefringence change), the positions of new spectral sidebands varied sensitively with the linear cavity birefringence. Cundiff et al interpreted these extra spectral sidebands as caused by the vector soliton polarization evolution in the cavity [7]. However, in our experiment it was found that even under the phase locked vector soliton case, such extra spectral sidebands could still appeared on the soliton spectrum.

To determine the physical origin of the extra sideband formation, we then conducted polarization resolved measurement of the vector soliton spectrum. To this end the laser output was first passed through a rotatable external cavity linear polarizer. To separate the two orthogonal polarization components of a vector soliton, we always first locate the orientation of the polarizer with maximum soliton transmission, which sets the long axis of an elliptically polarized vector soliton; we then rotate the polarizer by 90 degree to determine the soliton polarization component along the short axis of the polarization ellipse. Through separating the two orthogonal polarization components of the vector solitons, it turned out that the formation of the extra spectral sidebands was due to the coherent energy exchange between the two soliton polarization components. As can be clearly seen from the polarization resolved spectra, at the positions of extra spectral sidebands, while the spectral intensity of one soliton polarization component has a spectral peak, the orthogonal polarization component then has a spectral dip, indicting coherent energy exchange between them. We note that the energy flow between the two polarization components is not necessarily from the strong one to the weak one. Energy flow from the weak component to the strong component was also observed. In addition, it is to see from the polarization resolved spectra that the extra sidebands are symmetric with respect of the soliton peak frequency and at different wavelength positions the peak-dip can also alternate, suggesting that the energy exchange is the relative phase of the coupled components dependent. Fig. 2 (a) and (b) were obtained from the same laser but under different intra cavity polarization controller orientations. Obviously, the positions of the extra sidebands are the cavity birefringence dependent.

To verify our experimental observations and determine the extra sideband formation mechanism, we also numerically simulated the FWM in the laser. We used the following coupled Ginzburg-Landau equations to describe the pulse propagation in the weakly birefringent fibers in the cavity:

$$\begin{cases} \dfrac{\partial u}{\partial z} = i\beta u - \delta \dfrac{\partial u}{\partial t} - \dfrac{ik''}{2}\dfrac{\partial^2 u}{\partial t^2} + \dfrac{ik'''}{6}\dfrac{\partial^3 u}{\partial t^3} + i\gamma(|u|^2 + \dfrac{2}{3}|v|^2)u + \dfrac{i\gamma}{3}v^2 u^* + \dfrac{g}{2}u + \dfrac{g}{2\Omega_g^2}\dfrac{\partial^2 u}{\partial t^2} \\ \dfrac{\partial v}{\partial z} = -i\beta v + \delta \dfrac{\partial v}{\partial t} - \dfrac{ik''}{2}\dfrac{\partial^2 v}{\partial t^2} + \dfrac{ik'''}{6}\dfrac{\partial^3 v}{\partial t^3} + i\gamma(|v|^2 + \dfrac{2}{3}|u|^2)v + \dfrac{i\gamma}{3}u^2 v^* + \dfrac{g}{2}v + \dfrac{g}{2\Omega_g^2}\dfrac{\partial^2 v}{\partial t^2} \end{cases} \quad (1)$$

Where, u and v are the normalized envelopes of the optical pulses along the two orthogonal polarized modes of the optical fiber. $2\beta = 2\pi\Delta n/\lambda$ is the wave-number difference between the two modes. $2\delta = 2\beta\lambda/2\pi c$ is the inverse group velocity difference. k′ is the second order dispersion coefficient, k″ is the third order dispersion coefficient and represents the nonlinearity of the fiber. g is the saturable gain coefficient of the fiber and $\Omega_g$ is the bandwidth of the laser gain. For undoped fibers g=0; for erbium doped fiber, we considered its gain saturation as

$$g = G\exp\left[-\dfrac{\int(|u|^2 + |v|^2)dt}{P_{sat}}\right] \quad (2)$$

where G is the small signal gain coefficient and $P_{sat}$ is the normalized saturation energy.

The saturable absorption of the SESAM is described by the rate equation [8]:

$$\frac{\partial l_s}{\partial t} = -\frac{l_s - l_0}{T_{rec}} - \frac{|u|^2 + |v|^2}{E_{sat}} l_s \tag{3}$$

Where $T_{rec}$ is the absorption recovery time, $l_0$ is the initial absorption of the absorber, and $E_{sat}$ is the absorber saturation energy. To make the simulation possibly close to the experimental situation, we used the following parameters: $\gamma=3$ $W^{-1}km^{-1}$, $\Omega_g = 24$nm, $P_{sat}=100$ pJ, $k''_{SMF}=-23$ $ps^2/km$, $k''_{EDF}=-13$ $ps^2/km$, $k'''=-0.13$ $ps^3/km$, $E_{sat}=1$ pJ, $l_0=0.15$, and $T_{rec} = 6$ ps, Cavity length L= 10 m.

Fig. 3 shows the results obtained under different cavity birefringence. Extra spectral sidebands appeared clearly on the soliton spectrum. In particular, the extra sidebands of the orthogonal soliton polarization components exhibited out-of-phase variations. To verify that the extra spectral sidebands were caused by the FWM between the orthogonal soliton polarization components, we also deliberately removed the coherent coupling terms in our simulations. Without the FWM terms no extra spectral sidebands were observed. Numerically, it was also noticed that the appearance of the extra sidebands is related to the small linear cavity birefringence. When the linear cavity birefringence is set zero, although strong energy exchange exists between the two polarization components, no extra sidebands were observed, instead the overall soliton spectrum exhibits "peak-dip" alternation as the soliton propagates in cavity. Moreover, numerical simulations have also exhibited the dependence of extra sideband positions with the linear cavity birefringence.

The numerical simulations well reproduced the extra spectral sidebands and confirmed that their appearance is indeed caused by the FWM between the orthogonal soliton components. The result could also be easily understood. Due to small linear cavity birefringence, coherent coupling between the two polarization components of a vector soliton can no longer be neglected. Its existence causes coherent energy exchange between the two orthogonal soliton polarization components. Nevertheless, as far as the linear cavity birefringence is not zero, energy exchange does not occur at whole soliton spectrum, but only at certain wavelengths where the phase matching condition is fulfilled, which then leads to the formation of the discrete extra spectral sidebands.

In conclusion, we have experimentally observed extra spectral sideband generation on the soliton spectra of the phase locked vector solitons in a passively mode-locked fiber ring laser. Polarization resolved study on the soliton spectrum reveal that they are caused by the coherence energy exchange between the two orthogonal polarization components of the vector solitons. Numerical simulations have confirmed our experimental observation. Especially, numerical simulations show that FWM always exists under weak cavity birefringence. As far as the net cavity birefringence is not zero, phase matching condition can only be fulfilled at certain wavelengths. Our studies suggest that appearance of the sidebands is not a characteristic of the vector soliton polarization evolution, but the FWM between the components of a vector soliton.

**Figure captions:**

Fig. 1: Schematic of the experimental setup.

Fig. 2: Optical spectra of the phase locked vector solitons of the laser measured without passing and passing through a polarizer.

Fig. 3: Numerically calculated optical spectra of the vector solitons formed in fiber ring lasers.

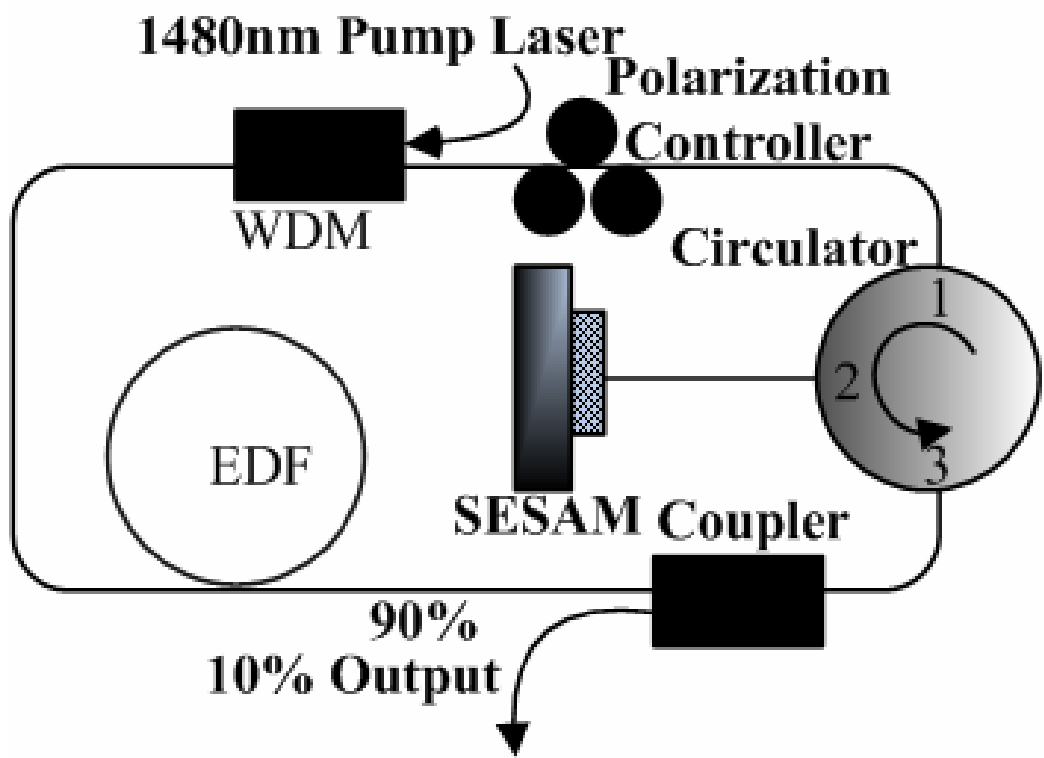

Fig. 1　　　H. Zhang et al.

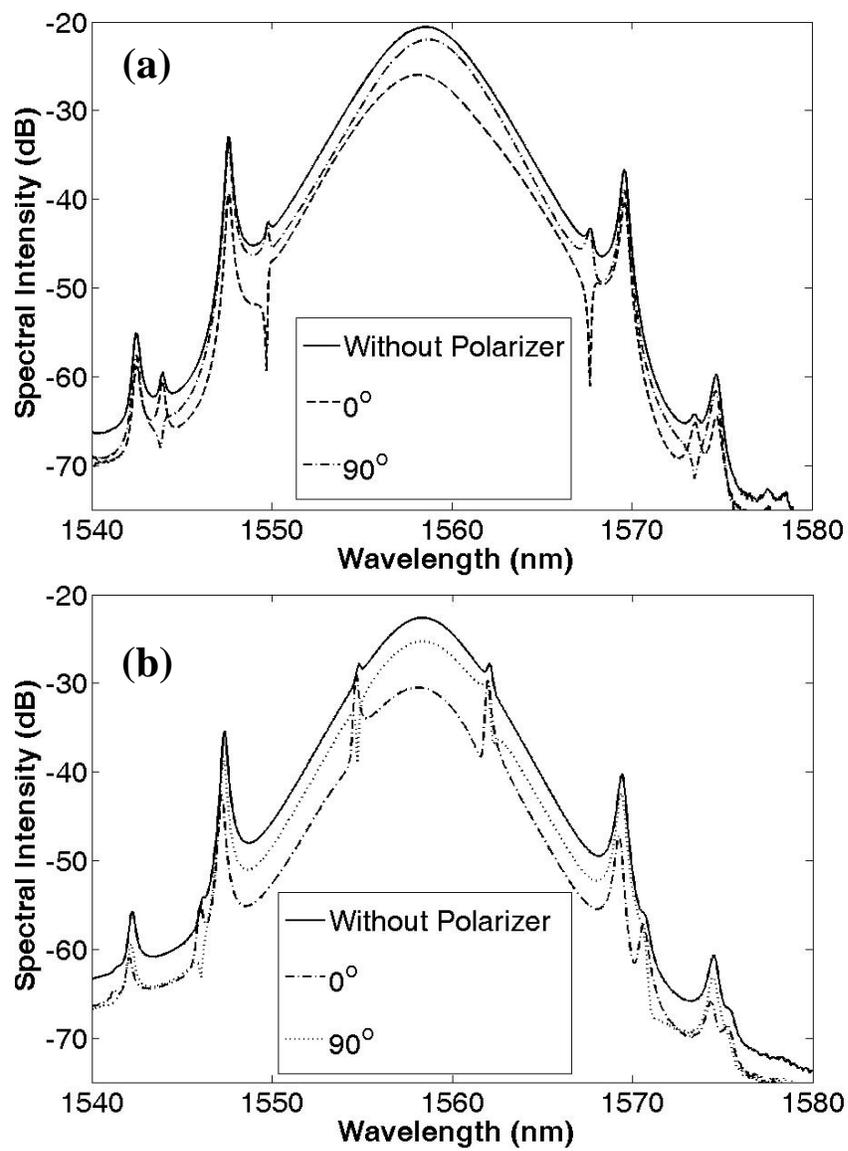

Fig. 2 H. Zhang et al.

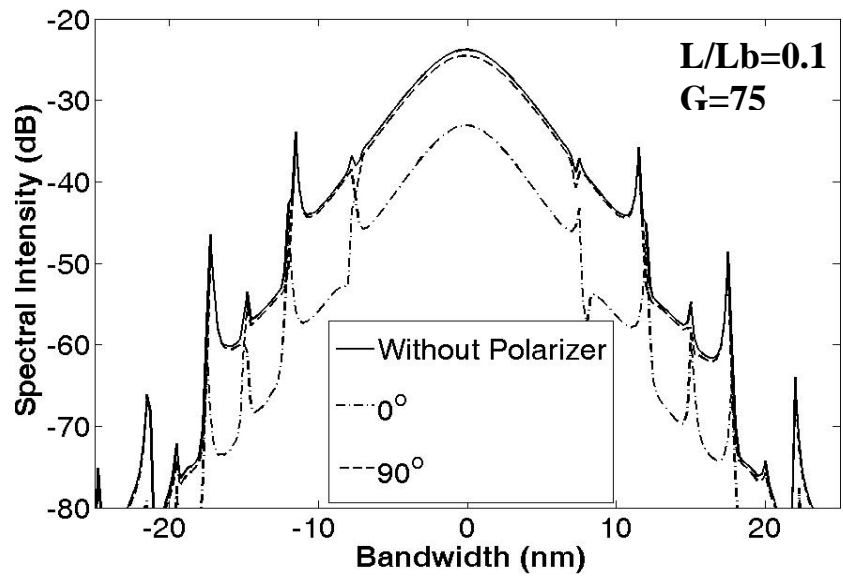

Fig. 3　　H. Zhang et al.